\documentclass{ifacconf}
\usepackage{amsmath} 
\usepackage{amsfonts}
\usepackage{amssymb}
\usepackage{subcaption}

\usepackage{enumerate}
\usepackage{algorithm}
\usepackage{algpseudocode}
\usepackage{amsmath}
\usepackage{xcolor}

\usepackage{tabularx}
\usepackage{booktabs}
\usepackage{array}
\usepackage[table]{xcolor}

\usepackage{booktabs}

\newtheorem{remark}{Remark}

\usepackage{bm}
\newcommand{\qedbox}{\rule{1.2ex}{1.2ex}}
    {\hfill\qedbox\par}

\usepackage{graphicx}      
\usepackage{natbib}        
\begin{document}
\begin{frontmatter}

\title{Distributed Traffic Signal Control of Interconnected Intersections: A Two-Lane Traffic Network Model\thanksref{footnoteinfo}} 

\thanks[footnoteinfo]{This research was supported in part by the National Natural Science Foundation of China under Grant 62122016; in part by LiaoNing Science and Technology Program under Grant 2023JH2/101700361 and Grant 2023JH2/101800024; and in part by the Fundamental Research Funds for the Central Universities under Grant DUT24ZD122. The work of Ting Bai and Andreas A. Malikopoulos was supported by MathWorks.}

\author[First]{Xinfeng Ru},
\author[Second]{Ting Bai},  
\author[First]{Weiguo Xia},  
\author[Second]{and Andreas A. Malikopoulos}

\address[First]{School of Control Science and Engineering, Dalian University of Technology, Dalian, China (e-mails: rxf@mail.dlut.edu.cn,   wgxiaseu@dlut.edu.cn). }
\address[Second]{Information and Decision Science Lab, School of Civil $\&$ Environmental Engineering, Cornell University, USA(e-mails: tingbai@cornell.edu, amaliko@cornell.edu).}

\begin{abstract}                
In this paper, we investigate traffic signal control in a network of interconnected intersections, aiming to balance lane-level vehicle densities through optimal green-time allocation. We develop a two-lane traffic flow model that explicitly captures lane-specific propagation dynamics, addressing key limitations of conventional road-level formulations. The proposed model offers a more granular and flexible representation of urban traffic, enabling controllers to react more accurately to lane-specific congestion patterns. Building on this model, we design a distributed model predictive control (MPC) framework and integrate it with the efficient alternating direction method of multipliers (ADMM) to enhance scalability and real-time performance. To accommodate time-varying traffic conditions, we further introduce a data-driven method for forecasting dynamic split ratios. Comprehensive VISSIM simulations on a six-intersection network in Dalian, China, demonstrate that the proposed approach outperforms existing signal control strategies in both traffic efficiency and computational speed, showing its promise for real-time deployment.
\end{abstract}

\end{frontmatter}

\section{Introduction}\label{Section 1}
Traffic congestion has become a pressing global challenge due to the rapid growth of vehicle populations and the limited expandability of existing infrastructure. Its consequences include long queues, inefficient road utilization, increased travel delays, and elevated environmental burdens. To mitigate traffic congestion, three major categories of control strategies have been explored: (i) signal offset control, which coordinates adjacent intersections to create green-wave progression~\citep{DeNunzio2015}, though its effectiveness declines under high traffic densities; (ii) route planning and optimization, which dynamically assigns vehicle routes~\citep{ba2015distributed,Bai2025routing}, but relies on advanced vehicle technologies and high user compliance~\citep{Li2025arouting}, restricting its practical applicability; and and (iii) green-time allocation, which adjusts signal phase durations to improve performance metrics such as vehicle density balancing~\citep{Grandinetti2018,ru2023distributed}. Among these strategies, green-time allocation emerges as a scalable and infrastructure-compatible solution with strong potential for network-wide congestion mitigation, which forming the central focus of this study. 

Traditional traffic models simplify traffic flow by treating each road as a single aggregated entity, merging the dynamics of multiple lanes into one state variable. Early fixed-time signal control methods employed historical traffic data to determine green times, with the objective of minimizing vehicle stops~\citep{Robertson1969}. Later, adaptive systems such as SCOOT~\citep{Robertson1991} and SCATS~\citep{Sims1980} incorporated real-time traffic measurements to adjust signal timings continuously. While these centralized approaches improve control performance, they often require extensive computational resources and communication, limiting their scalability and practicality in large urban networks.

To overcome the limitations of centralized approaches, distributed control strategies have emerged as practical alternatives for coordinating traffic across multiple interconnected intersections. For instance, a distributed model predictive control (MPC) scheme based on the store-and-forward model was proposed in~\citet{Camponogara2009} to balance intersection queues through green-time optimization. Using a similar idea, a road-based cell transmission model was developed in~\citet{Grandinetti2018}, where distributed averaging control was applied to regulate vehicle densities in a large-scale network. In addition, a model-free adaptive predictive control method was introduced in~\citet{ru2023distributed} to balance downstream densities without requiring a specific traffic model. Parallel to these efforts, max-pressure control~\citep{zaidi2016back} seeks to maximize intersection throughput; however, its reliance on idealized assumptions such as unlimited queue capacity restricts its effectiveness under real-world oversaturated conditions.

Recent advances in decentralized and distributed coordination of connected and automated vehicles (CAVs) have further highlighted the value of fine-grained modeling and predictive decision-making. \citet{Malikopoulos2020} established that CAVs can compute optimal trajectories using only local information, eliminating the need for centralized coordination. Building on this direction, \citet{chalaki2020TITS,chalaki2020TCST} proposed hierarchical and bilevel formulations that jointly optimize lane choice, arrival-time scheduling, and energy- or time-optimal control across adjacent intersections. A distributed mixed-integer quadratic programming scheme supported by a majorization–block-iterative optimization algorithm was later developed to integrate traffic signal control with vehicle-level coordination in mixed traffic~\citep{le2024distributed}, while \citet{tzortzoglou2024feasibility} analyzed feasibility limits in trajectory coordination and expanded the solution space via higher-order polynomial interpolation. Together, these studies underscore the value of lane-level modeling, predictive optimization, and distributed architectures in addressing the complexity of urban mobility systems.

\begin{figure} 
    \centering
    \includegraphics[scale=0.22]{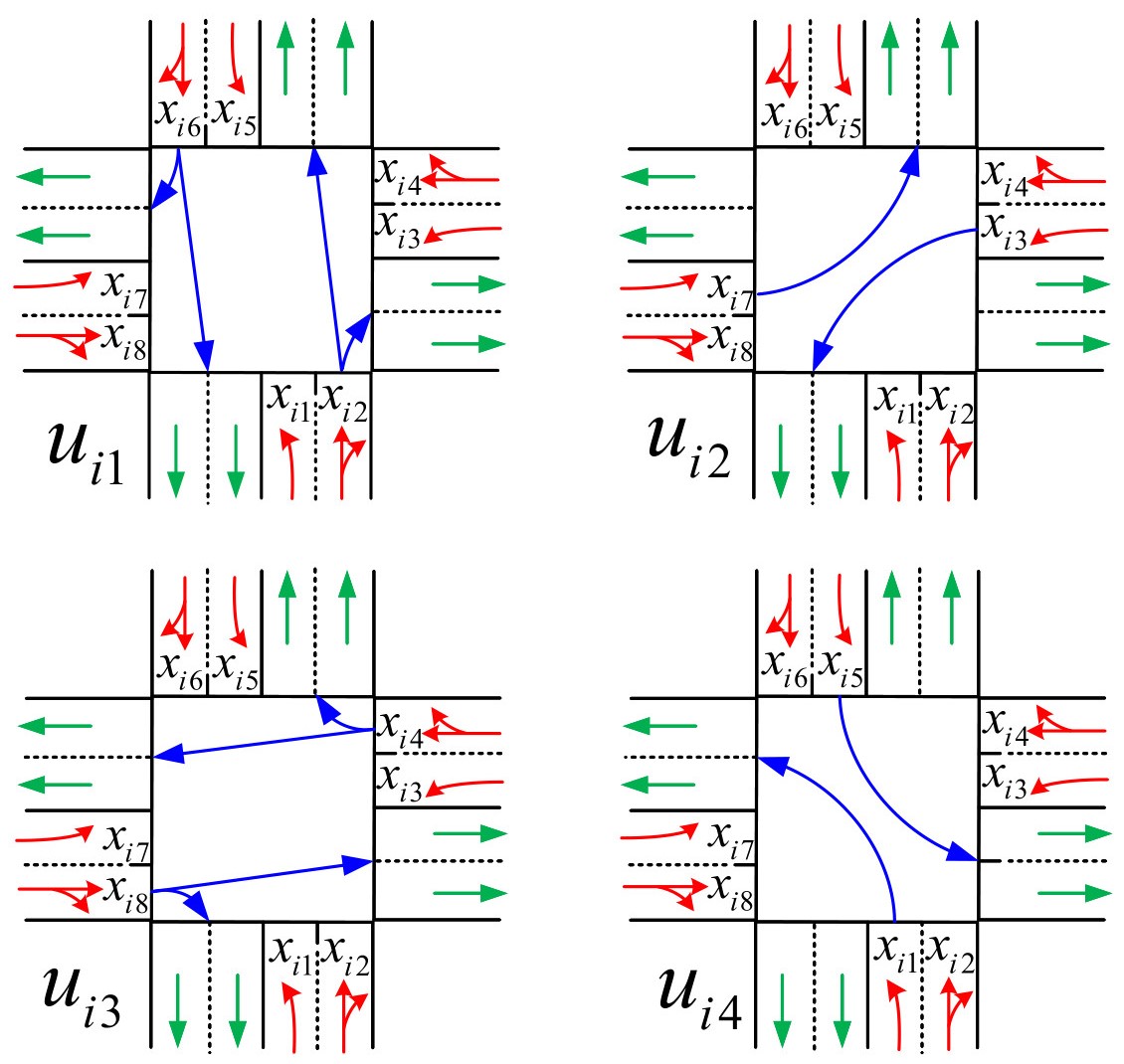}
    \vspace{-3pt}
    \caption{Illustration of a four-way intersection $i$, where lanes $x_{i2}$ and $x_{i6}$ are controlled by input $u_{i1}$; lanes $x_{i3}$ and $x_{i7}$ by $u_{i2}$; lanes $x_{i4}$ and $x_{i8}$ by $u_{i3}$; and lanes $x_{i1}$ and $x_{i5}$ by $u_{i4}$.}
    \label{Fig.1}
\end{figure}

Despite these achievements, most existing approaches still aggregate traffic at the road level, overlooking lane-specific dynamics. This simplification implicitly assumes that vehicles are evenly distributed across lanes---an assumption that is rarely valid in practice and often leads to inaccurate flow predictions. Although some studies introduce vehicle split ratios to approximate lane usage~\citep{de2010multi,yan2016extended}, these ratios are typically treated as fixed parameters and therefore cannot capture time-varying turning behaviors in dynamic environments. To address these limitations, we develop a two-lane traffic model that explicitly incorporates lane-level vehicle dynamics and turning movements. By modeling each road with two parallel lanes and tracking lane-specific vehicle distributions, the proposed model offers a more realistic description of traffic evolution. At each intersection, four signal phases govern lane-level vehicle movements, and their corresponding green times directly determine the discharge rates of each lane, as illustrated in Fig.~\ref{Fig.1}.

Building on the proposed two-lane model, we present a distributed MPC approach to allocate green times dynamically across a network of interconnected intersections. MPC is renowned for its ability to predict future system dynamics and avoid short-sighted decisions~\citep{8600793,ye2019survey}. The process of MPC typically involves three key steps: predicting system dynamics to anticipate future states, solving an optimization problem to compute optimal control actions, and implementing the actions over a rolling horizon to adapt to dynamic changes. To enhance the feasibility and scalability of the distributed MPC framework, we integrate the alternating direction method of multipliers (ADMM), which provides decomposition capabilities and fast convergence, making it well-suited for distributed optimization~\citep{boyd2011distributed,bai2020distributed}. In summary, this paper aims to achieve balanced vehicle density across an urban traffic network by introducing a two-lane traffic model and integrating distributed MPC with the ADMM algorithm. The main contributions are:  
\vspace{-2pt}
\begin{itemize}
\item In contrast to conventional road-based models, we propose a two-lane traffic model based on the store-and-forward framework, providing a more practical and accurate modeling of traffic flow propagation.
\vspace{2pt}

\item We present a distributed MPC approach for traffic signal control in an urban network of interconnected intersections. The efficient ADMM algorithm is further integrated to enhance real-time applicability and computational efficiency.
\vspace{2pt}

\item We develop a data-driven method to forecast time-varying split ratios, which enables the controller to adapt to dynamic turning movements that are often neglected in existing approaches.
\end{itemize}
Finally, a realistic simulation study in the VISSIM simulator is performed to evaluate the performance of the proposed method under realistic traffic conditions. Simulation results demonstrate the superior performance of the proposed MPC-ADMM controller, showing consistent improvements in average delay, number of stops, total travel time, and computation time in comparison with existing baseline methods.

The rest of this paper is organized as follows. Section~\ref{Section II} presents the two-lane traffic model and formulates the traffic signal control problem. Section~\ref{Section III} introduces the distributed MPC framework and its integration with the ADMM algorithm. Simulation results are provided in Section~\ref{Section IV}, and in Section~\ref{Section V}, we conclude the paper and discuss directions for future research.

\section{Problem Formulation}\label{Section II}
Consider a traffic network with $N$ interconnected intersections. Each intersection is linked to eight incoming roads, where vehicles travel in the same direction on each road, as illustrated in Fig.~\ref{Fig.1}. Each road contains two lanes: one dedicated to straight and right-turn movements, and the other to left-turn movements. To represent this network, we adopt a coupled-system framework in which each subsystem corresponds to an intersection and the four roads feeding into it. Specifically, the lanes marked by red arrows in Fig.~\ref{Fig.1} are associated with subsystem $i$, while those marked by green arrows belong to its neighboring subsystems. Each subsystem $i$ is described by a local state vector ${x_i}\!\in\!\mathbb{R}^n$ and a local control input ${u_i}\!\in\!\mathbb{R}^m$, where $n\!=\!8$ and $m\!=\!4$. The state vector ${x_i}\!=\![x_{i1}, \ldots, x_{i8}]^\top$ represents the number of vehicles in the eight incoming lanes of intersection $i$, while ${u_i}\!=\![u_{i1}, \ldots, u_{i4}]^\top$ specifies the green times allocated to the four traffic signal phases.

The interactions among the $N$ subsystems are modeled by a directed graph $\mathcal{G}\!=\!(\mathcal{V}, \mathcal{E})$, where the vertex set $\mathcal{V}\!=\!\{1, \ldots, N\}$ represents the set of subsystems and the edge set $\mathcal{E}\!\subseteq\!\mathcal{V}\!\times\! \mathcal{V}$ encodes their couplings. An edge $(j,i) \!\in\!\mathcal{E}$ with $i,j\!\in\!\mathcal{V}$ indicates that subsystem $j$ directly affects subsystem $i$ through a connecting road, making $j$ a neighbor of subsystem $i$. The set of neighbors of subsystem~$i$ is denoted by $\mathcal{N}_i$, with ${N}_i\!=\!|\mathcal{N}_i|$.

The vehicle outflow from subsystem $i$ is directly regulated by its local control input ${u_i}$. In contrast, the inflow into a subsystem is determined by the control inputs of its neighboring subsystems, resulting in coupled and interdependent dynamics across the network. The evolution of subsystem $i\!\in\!\mathcal{V}$ at time step $k \in \{0,1,\ldots\}$ is described as
\begin{align}
x_i(k\!+\!1) = x_i(k) - B_i(k)u_i(k) + C_i(k)z_i(k),\label{Equ.1}
\end{align}
where ${B_i}(k)\!\in\!\mathbb{R}^{n \times m}$ represents the outflow rate matrix, which can be obtained from road infrastructure measurements, and ${B_i}(k){u_i}(k)$ denotes the number of vehicles exiting subsystem $i$ at time step $k$. The term ${C_i}(k){z_i}(k)$ captures the inflow of vehicles contributed by neighboring subsystems. Here, ${C_i}(k) \in \mathbb{R}^{n \times m{N}_i}$ is the matrix of transfer rates, and ${z_i}(k) \in \mathbb{R}^{m{ N}_i}$ aggregates the control inputs of neighboring subsystems, given by
\begin{align}
z_i(k) = \left[ u_{j_1}^{\top}(k), \ldots, u_{j_{{ N}_i}}^{\top}(k) \right]^{\top},\label{Equ.2}
\end{align}
where $j_l\!\in\!{\cal N}_i$, $l\!=\!1,\dots,N_i$. The term ${C_i}(k){z_i}(k)$ is explicitly expressed as
\begin{align}
C_i(k)z_i(k) = \sum_{j_l \in {\cal N}_i} c_{j_li}(k) u_{j_l}(k),\label{Equ.3}
\end{align}
where $c_{j_li}(k)$, is the transfer rate from subsystem $j_l$ to subsystem $i$. Since the transfer rates $c_{j_li}(k)$ are influenced by multiple factors and directions, they are often difficult to measure directly and are treated as unknown parameters in this paper.

\begin{remark}
The traffic model proposed in~\citet{Bianchin2019} uses a single state variable to represent the number of vehicles on each road, which is then coupled with multiple control inputs. For example, as shown in Fig.~\ref{Fig.1}, combining the states $x_{i1}$ and $x_{i2}$ into a single state $\bar{x}_{i1}$ results in couplings with both control inputs $u_{i1}$ and $u_{i4}$. Different from their work, the two-lane traffic model described in~\eqref{Equ.1} decouples the state variables and control inputs, thus facilitating the integration of ADMM for efficient traffic signal optimization. 
\end{remark}

Unbalanced traffic flow distribution is a significant factor contributing to congestion, as it results in some lanes being overly congested while others remain underutilized. Inspired by \citet{ru2023distributed,Grandinetti2018}, the goal of this work is to achieve a balanced vehicle density distribution across lanes. Specifically, we aim to achieve a local uniform vehicle density distribution for each lane and its downstream connections by designing appropriate control inputs for each subsystem. To this end, a distributed MPC method is proposed to allocate green times for traffic signals, mitigating congestion caused by uneven traffic flow distribution. 

\section{ Distributed MPC for Traffic Signal Control via ADMM}\label{Section III}
This section presents a distributed MPC scheme to determine the green times of each signal phase across all intersections in the network. We start by reviewing the standard ADMM algorithm, followed by the formulation of the distributed MPC problem. Finally, we describe how ADMM is integrated into the MPC framework to improve computational efficiency. 

\subsection {Review of ADMM}
ADMM solves the following optimization problem:
\begin{align}
    \min \ \  & \ f(x)+g(z), \\
     \mathrm{s.\,t.} &  \  Ax + Bz =d,
\end{align}
 where $x \in \mathbb R^m, z\in \mathbb R^n$ are the control variables, $f\!:\!\mathbb R^m\!\to\! \mathbb R, ~g\!:\!\mathbb R^n\!\to\! \mathbb R$ are convex functions, and $A \in \mathbb R^{p\times m }, B \in \mathbb R^{p\times n }$ and $d \in \mathbb R^p$. ADMM makes use of the dual ascent method with the following augmented Lagrangian:
 \begin{align}
    {{\cal L}_\rho }(x,z,\lambda ) = & f(x) + g(z) + {\lambda ^{\top}}(Ax\! +\! Bz\! -\! d)   \nonumber \\ & + \frac{1}{2}\rho\left\| {Ax\! +\! Bz\! - \!d} \right\|_2^2,\nonumber
 \end{align}
 where $\lambda$ is the Lagrange multiplier, and $\rho$ is a positive constant
 to accelerate the convergence of the algorithm. In ADMM, the variables are updated by iteratively minimizing ${{\cal L}_\rho }(x,z,\lambda )$ with respect to $x$ and $z$. The iterations are given by
 \begin{align}
 \begin{cases}
 x^{k+1} = \arg \mathop {\min }\limits_x {{\cal L}_\rho }(x,z^k,\lambda^k ),\nonumber \\
 z^{k+1} = \arg \mathop {\min }\limits_z {{\cal L}_\rho }(x^{k+1},z,\lambda^k ),\nonumber \\
 \lambda^{k+1} = \lambda^{k} + \rho(Ax^{k+1}\!+\!Bz^{k+1}\!-\!d).\nonumber 
 \end{cases}
 \end{align}
To monitor the convergence of the algorithm towards optimality, the primal residuals $r^k$ and dual residuals $s^k$ of iteration $k$ are defined as
 \begin{align}
 r^{k+1} &= Ax^{k+1}+Bz^{k+1}-d,\nonumber \\
 s^{k+1} &= \rho A^{\top}B(z^{k+1}\!-\!z^{k}).\nonumber
 \end{align}
The stopping criterion is designed such that the primal residuals $r^k$ and dual residuals $s^k$ fall below the prescribed tolerance levels. In addition, a maximum iteration limit is imposed to ensure termination when an approximate solution is sufficient and desirable. We refer readers to~\citet{Boyd2011} for more details on the stopping criterion and convergence analysis of ADMM.
  
\subsection{Distributed MPC Framework}
Next, we present a distributed MPC framework for the coordinated signal control across interconnected intersections, which is grounded on the two-lane traffic model. By \eqref{Equ.1}, we derive the $M$-step ahead dynamic models below
\begin{align}
{x_i}(k\!+\!1) &= {x_i}(k) -{ B_i}(k){u_i}(k) + { C_i}(k){z_i}(k),\nonumber\\
{x_i}(k\!+\!2) &= {x_i}(k\!+\!1) - {B_i}(k\!+\!1){u_i}(k\!+\!1) \nonumber\\
    & \quad+ { C_i}(k\!+\!1){z_i}(k\!+\!1) \nonumber\\
	&= {x_i}(k) - { B_i}(k){u_i}(k) + { C_i}(k){z_i}(k) \nonumber\\
	&\quad - { B_i}(k\!+\!1){u_i}(k\!+\!1) + { C_i}(k\!+\!1){z_i}(k\!+\!1),\nonumber \\
	&\vdots \nonumber\\
	{x_i}(k\!+\!M) &= {x_i}(k) - \sum_{j = 0}^{M - 1} \!{B_i}(k\!+\!j){u_i}(k\!+\!j) \nonumber\\
	&\quad + \sum_{j = 0}^{M - 1}\!{C_i}(k\!+\!j){z_i}(k\!+\!j),\nonumber
\end{align}
which can be rewritten as the following compact form:
\begin{align}
&y_i(k\!+\!1) = \big[x_i^{\top}(k\!+\!1), \ldots , x_i^{\top}(k\!+\!M) \big]^{\top}, \nonumber\\
&\mathcal B_i(k) = 
\begin{bmatrix}
B_i(k)      & \mathbf{0}   & \cdots & \mathbf{0} \\[2pt]
B_i(k)      & B_i(k\!+\!1)     & \cdots & \mathbf{0} \\[2pt]
\vdots      & \vdots       & \ddots & \vdots     \\[2pt]
B_i(k)      & B_i(k\!+\!1)     & \cdots & B_i(k\!+\!M\!-\!1)
\end{bmatrix}, \nonumber\\
&\mathcal C_i(k) = 
\begin{bmatrix}
C_i(k)      & \mathbf{0}   & \cdots & \mathbf{0} \\[2pt]
C_i(k)      & C_i(k\!+\!1)     & \cdots & \mathbf{0} \\[2pt]
\vdots      & \vdots       & \ddots & \vdots     \\[2pt]
C_i(k)      & C_i(k\!+\!1)     & \cdots & C_i(k\!+\!M\!-\!1)
\end{bmatrix}, \nonumber \\
&U_i(k) = \big[ u_i^{\top}(k), \ldots , u_i^{\top}(k\!+\!M\!-\!1) \big]^{\top}, \nonumber \\
&Z_i(k) = \big[ z_i^{\top}(k), \ldots , z_i^{\top}(k\!+\!M\!-\!1) \big]^{\top}, \nonumber \\
&E = \mathbf{1}_M,\nonumber
\end{align}
where $\bm {1}_M$ denotes the $M$-dimensional column vector of all ones. Note that the matrix ${{\mathcal C}_i}(k)$ is unknown, and its entry ${C_i}(k)$  can be estimated by minimizing the following cost function~\citep{Hou2016}:
\begin{align}
	\!\!\!J({C _i}(k))\!= & \big\|{x_i}(k)\!-\!{x_i}(k\!-\!1)\!+\!{ B_i}(k\!-\!1){u_i}(k\!-\!1)\nonumber \\ &-\!{C_i}(k){z_i}(k\!-\!1)\big\|_2^2\!+\!{\mu _i}\big\|{{C_i}(k)\!-\!{{\hat C}_i}(k\!-\!1)} \big\|_2^2,\label{Equ.6}
\end{align}
where ${{\hat C }_i}(k\!-\!1)$ denotes the estimate of ${C _i}(k\!-\!1)$, $\mu_i\!>\!0$ is a weighting factor to restrain the exaggerated change of pseudogradients, and ${\left\| x \right\|_2}$ denotes the 2-norm of vector~$x$. Accordingly, the estimate ${{\hat { C}}_i}(k)$ can be updated by minimizing~\eqref{Equ.6} with respect to ${C_i}(k)$ as
\begin{align}\hspace*{-0cm}
{{\hat C}_i}(k)\!= & {{\hat C}_i}(k\!-\!1)\!+\!\big[{x_i}(k) \!- \!{x_i}(k\! - \!1) \!+ \!{B_i}(k\!-\!1){u_i}(k\!-\!1) \nonumber \\
& - {{\hat C}_i}(k\!-\!1){z_i}(k\!-\!1)\big]{z_i}^{\top}(k\!-\!1) \nonumber \\
& \times {\big[{\mu _i}I_{{ m{ N_i}}}\!+\!{z_i}(k\!-\!1){z_i}^{\top}(k\!-\! 1)\big]^{ - 1}},\label{Equ.7}
\end{align}
where  $I_{{ m{ N_i}}}$ denotes the ${{ m{ N_i}}}$ dimensional identity matrix. The estimation method introduced above cannot be applied to obtain the estimates of ${B _i(k\!+\!1)}, \ldots, {B _i(k\!+\!M\!-\!1)}$ in ${\mathcal B_i}(k)$ or the estimates of ${C _i(k\!+\!1)}, \ldots, {C_i(k\!+\!M\!-\!1)}$ in ${\mathcal C_i}(k)$ directly. Since ${\mathcal  B_i}(k)$ and ${\mathcal C_i}(k)$ contain many zero entries, we append the nonzero elements of matrices ${B _i(k\!+\!1)}, \ldots,$ ${B _i(k\!+\!M\!-\!1)}$ into ${ b _i(k\!+\!1)}, \dots, { b _i(k\!+\!M\!-\!1)}$, and similarly collect the nonzero elements of ${C _i(k\!+\!1)}, \dots,$ ${C _i(k\!+\!M\!-\!1)}$ into ${ c _i(k\!+\!1)}, \dots,{ c _i(k\!+\!M\!-\!1)}$. A multi-layer hierarchical forecasting method is then employed to forecast these variables, and we obtain
\begin{align}
{{\hat { b} }_i}(k\!+\!j) &=  {\varphi_1}(k){{\hat { b} }_i}(k\!+\!j\!-\!1) + {\varphi _2}(k){{\hat {b} }_i}(k\!+\!j\!-\!2)\nonumber \\ 
&\ \ \ +  \cdots  + {\varphi _p}(k){{\hat { b} }_i}(k\!+\!j\!-\!p), \nonumber\\  
{{\hat {c}  }_i}(k\!+\!j) &= {\theta _1}(k){{\hat { c} }_i}(k\!+\!j\!-\!1) + {\theta _2}(k){{\hat { c} }_i}(k\!+\!j\!-\!2)\nonumber \\ 
& \ \ \ +  \cdots  + {\theta _p}(k){{\hat {c} }_i}(k\!+\!j\!-\!p),\nonumber
\end{align}
where ${{\hat { b}}_i}(k), {{\hat {c}}_i}(k)$ denote the estimates of ${ b_i}(k)$ and ${c_i}(k)$, respectively, with $j\!=\!1,\ldots, M\!-\!1$, and $p$ is an appropriate order which is normally set as 2-7~\citep{Li2021,Hou2013}. Then, let us define
\begin{align}
{\varphi (k)}&\buildrel \Delta \over =  \big[ {{\varphi _1}(k), \ldots ,{\varphi _p}(k)} \big]^{\top},\nonumber\\
{\theta (k)}&\buildrel \Delta \over = \big[ {{\theta _1}(k), \ldots ,{\theta _p}(k)} \big]^{\top},\nonumber
\end{align}
which can be updated as~\citep{Li2021} 
\begin{align}
{\varphi (k)} &= {\varphi (k\!-\!1)} + \frac{{\hat \eta _i(k\!-\!1)}}{{\delta + {{\left\| {{{\hat \eta }_i}(k\!-\! 1)} \right\|}_2}}} \nonumber \\ 
& \ \ \ \times \!\left[ {{{\hat {b} }_i}(k) - {{\hat \eta }_i}^{\top}(k\!-\!1){\varphi (k\!-\!1)}} \right],\label{Equ.8}\\
{\theta (k)} & = {\theta (k\!-\!1)} + \frac{{\hat \Xi _i(k\!-\!1)}}{{\delta + {{\left\| {{{\hat \Xi }_i}(k\!-\!1)} \right\|}_2}}}  \nonumber \\ 
& \ \ \ \times\!\big[ {{{\hat {c} }_i}(k) - {{\hat \Xi }_i}^{\top}(k\!-\!1){\theta (k\!-\!1)}} \big],\label{Equ.9}
\end{align}
where the terms 
\begin{align}
{{\hat \eta }_i}(k\!-\!1)\!=\!\big[ {{{\hat { b} }_i}(k\!-\!1), \ldots ,{{\hat {b} }_i}(k\!-\!p)} \big]^{\top},\nonumber\\
{{\hat \Xi }_i}(k\!-\!1)\!=\!\big[ {{{\hat { c} }_i}(k\!-\!1), \ldots ,{{\hat { c} }_i}(k\!-\!p)} \big]^{\top},\nonumber
\end{align}
and $\delta \!\in\!\left( {0,1} \right]$ is designed to avoid that the denominator equals zero.

\subsection{Distributed Traffic Signal Control via ADMM }
A coordinated and scalable solution to the distributed MPC problem formulated in the previous subsection is achieved by employing ADMM, which decomposes the global objective into local subproblems that can be solved in parallel.  To achieve a locally uniform vehicle density distribution for each lane relative to its downstream links~\citep{ru2023distributed,Grandinetti2018}, we formulate the network-level cost function as the following form:
\begin{subequations}
\begin{align}\hspace*{-0.1cm}
\min \sum\limits_{i = 1}^N \phi_i(k) 
\!=\! & \sum\limits_{i = 1}^N \!\Bigg( \sum\limits_{m = 1}^8 \sum\limits_{h = 1}^M 
\Big( \rho_{im}(k\!+\!h) - \bar\rho_{im}(k\!+\!h) \Big)^2 \nonumber \\
& \quad \quad\quad  \sum\limits_{h = 1}^{M-1}\!u_i^{\top}(k\!+\!h) R_i u_i(k\!+\!h) \Bigg), \label{VM}
\end{align}
subject to
\begin{align}
& x_i(k\!+\!1) = x_i(k)\!-\!B_i(k) u_i(k)\!+\!C_i(k) z_i(k), \label{st1} \\
& \bar \rho_{im}(k\!+\!h) = \frac{1}{|N_{im}^+|} 
\!\!\!\sum\limits_{jg \in N_{im}^+} \!\!\!\!\frac{x_{jg}(k\!+\!h)}{L_{jg}}, 
\quad h\!=\!1,\ldots,M, \\
& \sum\limits_{m = 1}^4\!u_{im}(k\!+\!h)\!+\!Q_i = S, 
\quad h\!=\!1,\ldots,M\!-\!1, \\
& u_{im}(k\!+\!h) \in [u_{\min}, u_{\max}], 
\quad h\!=\!1,\ldots,M\!-\!1, \label{st2}
\end{align}
\end{subequations}
where $x_{im}(k)$ and $u_{im}(k)$ denote the $m$th components of $x_i(k)$ and $u_i(k)$, respectively. 
The term $\bar\rho_{im}(k\!+\!h)$ represents the average density of the downstream lanes of lane $im$ at the prediction step $h$, with $L_{im}$ (and $L_{jg}$) denoting the lane length, 
and $N_{im}^+$ denoting the set of its downstream lanes of cardinality $|N_{im}^+|$. 
The weighting matrix $R_i\!=\!\mathrm{diag}(r_{i1},r_{i2},r_{i3},r_{i4})$ is diagonal and positive definite, ensuring proper scaling of the two terms {in~\eqref{VM}.} 
Moreover, $Q_i$ denotes the yellow-light duration at subsystem $i$, 
and $S$ is the common cycle length across all intersections. 
The parameters $u_{\min}$ and $u_{\max}$ specify the minimum and maximum green times, respectively. The first term {in~\eqref{VM}} penalizes the deviation of lane densities from their downstream averages, while the second term regularizes control efforts to enhance convergence.

By solving the optimization problem (\ref{VM}) with constraints (\ref{st1})-(\ref{st2}), one can derive the green time of each phase of all the intersections with the optimal performance for the entire network. However, the centralized optimization method involves massive data transmission and has high online computational complexity, which is impractical for multiple interconnected intersections. From the above analysis, an intersection belongs to only one subsystem after the traffic system decomposition. Hence, a distributed optimization framework for urban traffic signals is proposed. The optimization problem for each subsystem $i\!=\!1,\dots,N$ is presented as
\begin{align}\label{Vp}
\min {\phi _i}(k) = & \sum\limits_{m = 1}^8 \sum\limits_{h = 1}^M \Big( {\rho _{im}}(k\!+\!h) - {{\bar \rho }_{im}}(k\!+\!h) \Big)^2 \nonumber \\
& + \sum\limits_{h = 1}^{M - 1}\! {u_i}^{\top}(k){R_i}{u_i}(k),\\
\mathrm{s.\,t.}  \ \ &	\textnormal{(\ref{st1})-(\ref{st2})}.\nonumber
\end{align}
Although the optimization problem of the whole network in~\eqref{VM} can be decomposed into a distributed optimization problem in~(\ref{Vp}), it remains difficult to implement in large-scale urban networks, since solving~(\ref{Vp}) with constraints~(\ref{st1})--(\ref{st2}) directly may not satisfy real-time traffic control requirements~\citep{Xin2011,farokhi2013distributed}. Note that, in our proposed two-lane traffic model, the state variables and control inputs are decoupled, which allows the optimization problem~\eqref{Vp} to be further split into independent subproblems. Accordingly, the distributed MPC problem can be rewritten as  
\begin{align}
   \min {\phi _i}(k) = \sum_{m=1}^4 f_{im}\big(U_{im}(k)\big), \quad i\!=\!1,\ldots,N,\label{Ve}
\end{align}
subject to \eqref{st1}--\eqref{st2}, where
\begin{align}
f_{im}(U_{im}(k)) &= \sum_{h=1}^M \sum_{j \in \mathcal{L}_{im}} 
\!\!\Big(\rho_{ij}(k\!+\!h) - \bar\rho_{ij}(k\!+\!h)\Big)^2  \nonumber\\
 &\quad + U_{im}^{\top}(k)\, r_{im}\, U_{im}(k),
\end{align}
with $U_{im}(k)\!=\!\big[u_{im}^{\top}(k), \ldots, u_{im}^{\top}(k\!+\!M\!-\!1)\big]^{\top}$, 
$r_{im}$ denoting the $m$th diagonal element of $R_i$, and $\mathcal{L}_{im}$ being the set of lanes associated with control group $m$.  

Since the ADMM provides strong convergence properties and is well-suited for decomposable optimization problems with constraints, it is employed to {solve~\eqref{Ve}} efficiently at each time step $k$. The detailed procedure is summarized in {Algorithm~1}. At every time step, each subsystem solves its optimization problem subject to constraints (\ref{st1})-(\ref{st2}) after receiving the related information from its neighboring subsystems. The subsystem then shares the intermediate solution with its neighboring subsystems to enable solving their optimization problems. Finally, all the control inputs $U_i(k), i\!=\!1,\ldots,N$ of the entire network can be obtained, and each subsystem $i$ applies the first control of $U_i(k)$ as the optimal control.
\begin{algorithm}[t]
\caption{Distributed Signal Control via ADMM}
\label{alg:admm}
\begin{algorithmic}[1]

\Statex \textbf{Input:} Maximum computation time $T_{\max}$; stopping threshold $\varepsilon_{\mathrm{stop}}$.
\Statex \textbf{Output:} Control inputs $U_{i1}(k),U_{i2}(k),U_{i3}(k),U_{i4}(k)$.

\For{$i = 1,2,\dots,N$ \textbf{ in parallel}}
    \State Initialize $s \gets 1$, $t_i(k)\gets 0$, and measure $x_i(k)$.

    \While{$\varepsilon^{(s)} \ge \varepsilon_{\mathrm{stop}}$ \textbf{ and } $t_i(k)\le T_{\max}$}
        \State Receive $U_j(k\!-\!1)$ and $y_j(k\!-\!1)$ from neighbors.
        \State Update $\mathcal{B}_i(k)$ and $\mathcal{C}_i(k)$ by \eqref{Equ.6}--\eqref{Equ.9}.\State Solve the optimization problem
						\begin{align}
							&\min {\cal L}_{\rho }(U_{i1}(k),U_{i2}(k),U_{i3}(k),U_{i4}(k),\lambda) \nonumber\\ 
							&= \sum\limits_{m = 1}^4 \!{f_{im}}({U_{im}}(k)) \!+\! \frac{\rho}{2}{\|\vartheta (k)\|}^2 \!+\! {\lambda}^{\top} \vartheta (k), \notag
						\end{align}
 \Statex \hspace{1cm}where 

\Statex \hspace{1cm}$%
\vartheta(k)=
\begin{bmatrix}
    \sum_{m=1}^4 u_{im}(k\!+\!1) + Q_i - S \\
    \sum_{m=1}^4 u_{im}(k\!+\!2) + Q_i - S \\
    \vdots \\
    \sum_{m=1}^4 u_{im}(k\!+\!M\!-\!1) + Q_i - S
\end{bmatrix},$

\Statex \hspace{1cm}$%
\lambda =
\begin{bmatrix}
    \lambda_1,\ldots,\lambda_{M-1}
\end{bmatrix}^{\!\top}. $

        \State \normalsize{\textbf{Block updates:}}
        \State \small{$U_{i1}^{(s)} \!\!\gets\! \displaystyle\arg\min_{U_{i1}}
                \mathcal{L}_\rho\big(U_{i1},U_{i2}^{(s-1)},U_{i3}^{(s-1)},U_{i4}^{(s-1)},\lambda^{(s-1)}\big)$,}
        \State $U_{i2}^{(s)}\!\! \gets\! \displaystyle\arg\min_{U_{i2}}
                \mathcal{L}_\rho\big(U_{i1}^{(s)},U_{i2},U_{i3}^{(s-1)},U_{i4}^{(s-1)},\lambda^{(s-1)}\big),$
        \State $U_{i3}^{(s)} \!\!\gets\! \displaystyle\arg\min_{U_{i3}}
                \mathcal{L}_\rho\big(U_{i1}^{(s)},U_{i2}^{(s)},U_{i3},U_{i4}^{(s-1)},\lambda^{(s-1)}\big),$
        \State $U_{i4}^{(s)} \!\!\gets\! \displaystyle\arg\min_{U_{i4}}
                \mathcal{L}_\rho\big(U_{i1}^{(s)},U_{i2}^{(s)},U_{i3}^{(s)},U_{i4},\lambda^{(s-1)}\big),$

        \State $\lambda^{(s)} \!\!\gets \lambda^{(s-1)} +
               \rho \cdot \vartheta\big(U_{i1}^{(s)},U_{i2}^{(s)},U_{i3}^{(s)},U_{i4}^{(s)}\big).$

        \State \normalsize{Update $t_i(k)$.}
        \State \normalsize{$\varepsilon^{(s+1)} \!\gets\!
               \big\| \lambda^{(s)}\!-\!
               \vartheta\big(U_{i1}^{(s)},U_{i2}^{(s)},U_{i3}^{(s)},U_{i4}^{(s)}\big)
               \big\|_\infty$}.

        \State $s \!\gets\! s\!+\!1$; send $U_i(k)$ and $y_i(k)$ to neighbors.
    \EndWhile

\EndFor

\end{algorithmic}
\end{algorithm}

\section{Simulation Study}\label{Section IV}
In this section, we conduct a realistic simulation study to evaluate the performance of the proposed two-lane traffic model and the distributed MPC scheme. We first outline the simulation setup and then introduce the benchmark control methods, as well as the performance metrics used for evaluation. Finally, the section concludes with a comparative analysis of simulation results.
\subsection{Simulation Setup}
\begin{figure}[t]
    \centering
    \begin{subfigure}[b]{0.45\textwidth}
        \centering
        \includegraphics[scale=0.24]{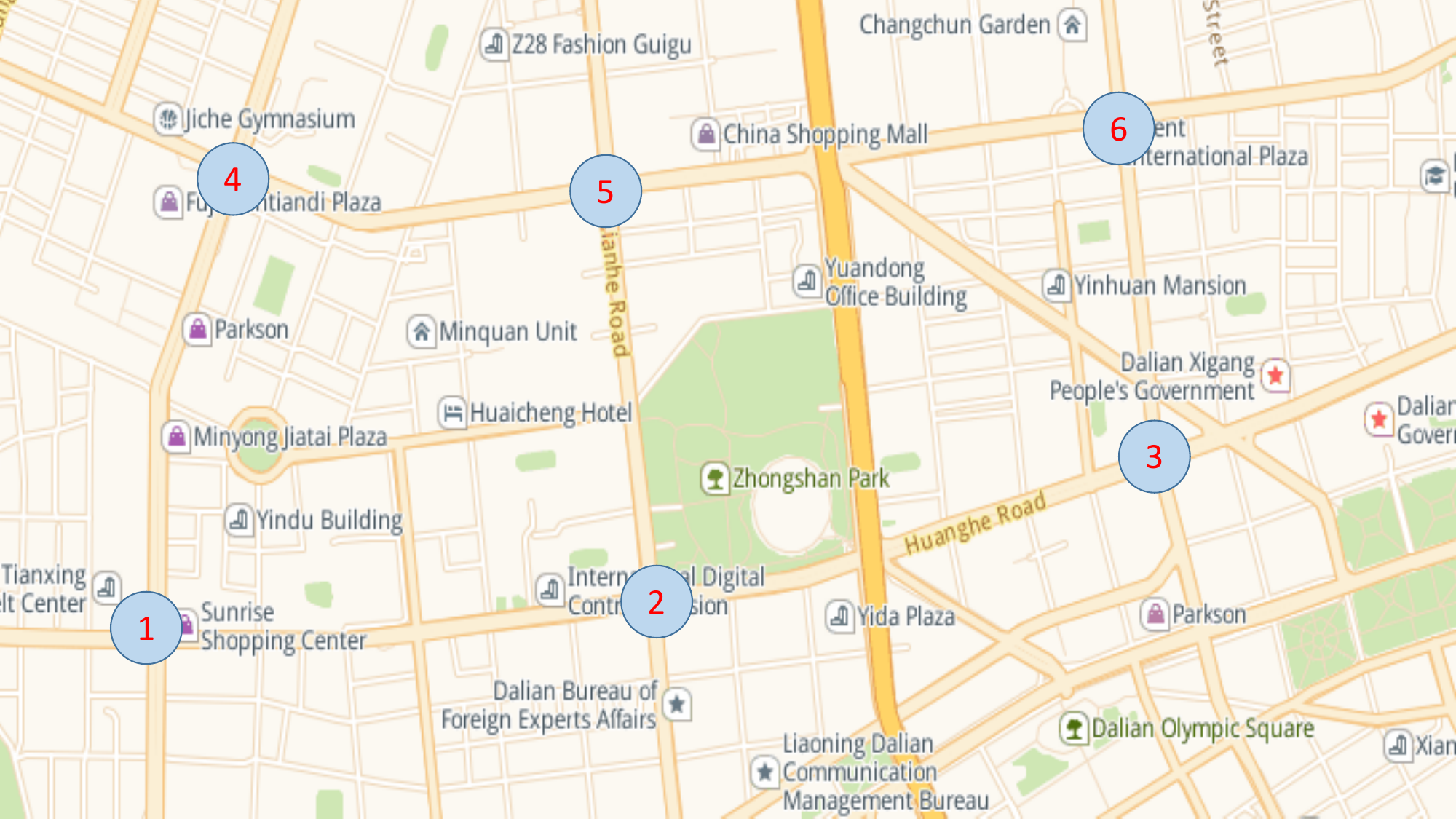}
        \caption{Geographic map of the selected intersections.}
        \label{fig:network_a}
    \end{subfigure}
    \hfill
    \begin{subfigure}[b]{0.45\textwidth}
        \centering
        \includegraphics[scale=1.5]{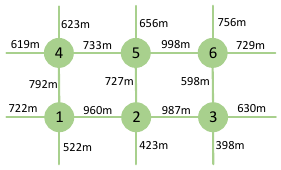}
        \caption{Topological representation of the traffic network.}
        \label{fig:network_b}
    \end{subfigure}
    \caption{Part of a traffic network of Dalian City, China.}
    \label{fig:network}
\end{figure}

The effectiveness of the proposed two-lane traffic model and distributed MPC scheme is evaluated through simulations on a six-intersection traffic network in Dalian, China, comprising 68 roads, as illustrated in Fig.~\ref{fig:network}. The traffic environment is simulated in VISSIM, while the control algorithms are implemented in Python. The simulation settings are as follows: the sampling period and the common cycle length of all intersections are both fixed at $S\!=\!120$ seconds. The total simulation horizon is 7200 seconds, corresponding to 60 control intervals. The traffic demand is designed to replicate morning rush-hour conditions in Dalian, ranging from 300 to 800 vehicles per hour {per road}. For each traffic phase, the green time is constrained within $u_{\min}\!=\!10$ seconds and $u_{\max}\!=\!70$ seconds. In addition, the prediction horizon of the MPC controller is set to $M\!=\!5$, meaning that five future time steps are considered in the optimization process.

\subsection{Comparison with Baseline Methods}
The performance of the proposed distributed MPC scheme is evaluated through comprehensive comparisons with the following baseline control strategies:
\vspace{-3pt}
\begin{enumerate}[1)]
    \item [$\bullet$] \textbf{Fixed-Time Control (FTC)}: Each intersection operates under a fixed signal timing plan as proposed in~\citep{Robertson1969}, where every signal phase is allocated one-quarter of the total cycle time. This strategy serves as a baseline to demonstrate the limitations of conventional fixed-time traffic control.
    \vspace{2pt}
    
    \item [$\bullet$] \textbf{Max-Pressure Control}: A widely adopted real-time adaptive strategy that selects signal phases by maximizing the pressure difference between upstream and downstream lanes. This method is effective in mitigating congestion under various traffic conditions and serves as a strong model-free benchmark.
    \vspace{2pt}

    \item [$\bullet$] \textbf{Model-Free Adaptive Predictive Control \\ (MFAPC)}: A distributed adaptive predictive control method developed in \citep{ru2023distributed}, which updates the pseudogradient online without requiring an explicit traffic model. This baseline evaluates the benefits of incorporating the proposed two-lane model and ADMM over existing data-driven control schemes.
    \vspace{2pt}

    \item [$\bullet$] \textbf{MPC-Road}: To assess the advantages of the proposed two-lane traffic model, we also compare it with a road-level distributed MPC scheme introduced in~\citep{Camponogara2009}. In this method, the green time for each phase is computed based on a single-road store-and-forward model, while not capturing the lane-level dynamics.
    \vspace{2pt}

    \item [$\bullet$] \textbf{MPC-CasADi}: The distributed MPC problem is implemented in the CasADi optimization framework using its default nonlinear programming solver. This implementation serves as a baseline for evaluating the computational efficiency of the proposed ADMM-based MPC scheme.
\end{enumerate}

\subsection{Performance Evaluation Metrics}
\begin{table*}[t]
    \centering
    \footnotesize
    \caption{Performance metrics under different control strategies.}
    \label{tab:results}
    \begin{tabularx}{\textwidth}{l *{4}{>{\centering\arraybackslash}X}}
        \toprule
        \textbf{Control Strategy} & \textbf{Average Delay (sec)} & \textbf{Average Stops} & \textbf{Total Travel Time (min)} & \textbf{Average Computation Time (sec)}  \\
        \midrule
        \rowcolor{blue!10}
        Proposed Method     & 105.164 & 3.129 & 937.502 & 1.323 \\
        MPC-CasADi   & 106.468 & 3.154 & 941.562 & 4.137 \\
        MFAPC         & 132.854 & 4.210 & 1018.436 & 0.972 \\
        Max-Pressure   & 148.372 & 4.856 & 1049.273 & /     \\
        MPC-Road      & 164.235 & 5.264 & 1089.354 & 4.389 \\
        FTC           & 229.500 & 8.858 & 1345.288 & /     \\
        \bottomrule
    \end{tabularx}
\end{table*}

The performance of different control strategies is evaluated using three key performance indicators: average density, average flow rate, and relative loss time. Specifically,  
\vspace{-3pt}
\begin{itemize}
\item [1)]\textit{Average Density:} It represents the mean vehicle density across all roads in the network. A lower value indicates less congestion and higher throughput, reflecting improved network performance.
\vspace{2pt}

\item [2)]\textit{Average Flow Rate:} It characterizes the overall traffic condition of the network. It captures multiple aspects such as vehicle delays, travel speeds, and network throughput. Higher flow rates correspond to smoother traffic flow and reduced congestion.
\vspace{2pt}

\item [3)]\textit{Relative Loss Time:} It measures the additional travel time experienced by vehicles compared to free-flow conditions. Larger values indicate more severe congestion and greater travel delays.
\end{itemize}

In addition to these three primary indicators, we also report several supplementary metrics, including the average travel time delay, average number of stops, total travel time, and average computation time. These indicators provide complementary insights into driving comfort, network-wide efficiency, and the real-time feasibility of different control strategies.

\subsection{Simulation Results and Analysis}
\begin{figure}[t] 
	\centering
	\includegraphics[width=8.6cm, height=5.6cm]{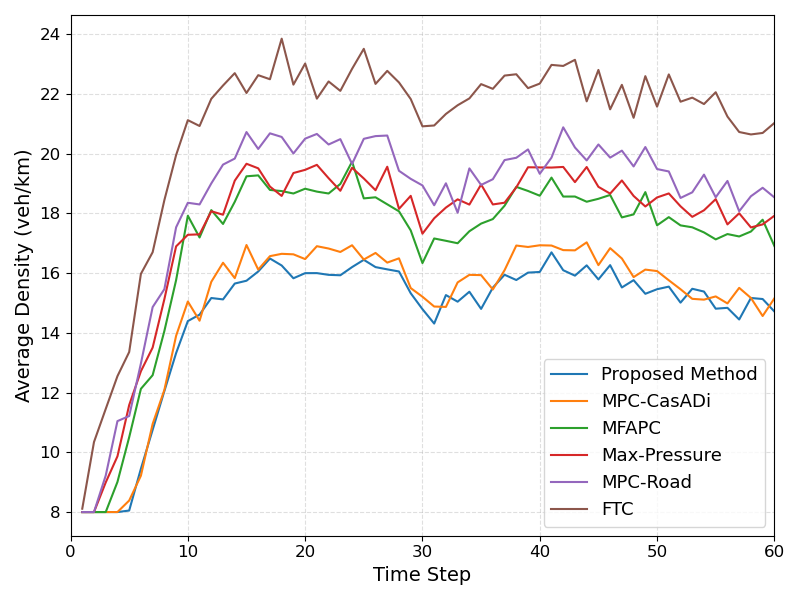}
	\caption{Comparison of average vehicle density.}
	\label{fig:density}
\end{figure}

\begin{figure}[t] 
	\centering
	\includegraphics[width=8.6cm, height=5.6cm]{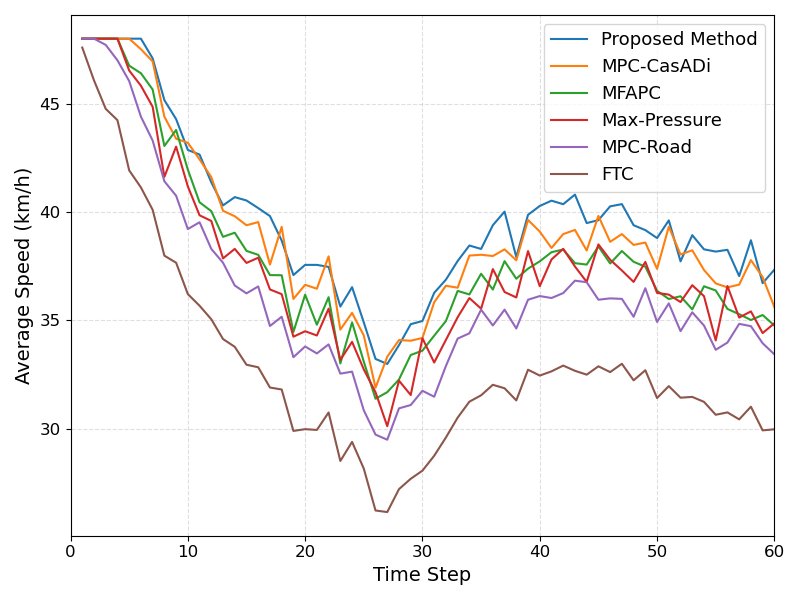}
	\caption{Average vehicle speed in different control strategies.}
	\label{fig:speed}
\end{figure}
 
\begin{figure}[t] 
	\centering
	\includegraphics[width=8.6cm, height=5.6cm]{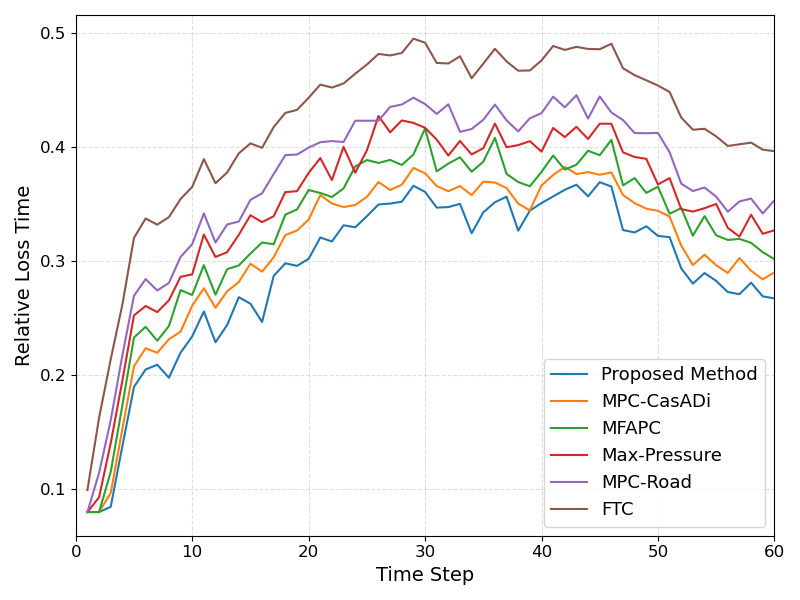}
	\caption{Relative loss time in different control strategies.}
	\label{fig:RLT}
\end{figure}

The comparison results are illustrated in Figs.~\ref{fig:density}--\ref{fig:RLT} and summarized in Table~\ref{tab:results}. Specifically, Fig.~\ref{fig:density} presents the evolution of the average density on the traffic network, Fig.~\ref{fig:speed} depicts the network-wide average vehicle speed, and Fig.~\ref{fig:RLT} shows the relative loss time under each control strategy. The results show that the proposed method achieves lower per-lane average density, higher network-level speed, and smaller relative loss time compared with the other benchmark approaches. In addition, Table~\ref{tab:results} shows that the proposed method achieves the best overall performance, yielding lower delays, fewer stops, shorter travel times, and reduced computation burden.

From these results, it is evident that the proposed MPC-ADMM strategy achieves the best overall performance across all evaluated metrics. It yields the lowest average delay, the fewest stops, the shortest total travel time, and the fastest computation time, demonstrating both strong control effectiveness and excellent real-time feasibility. Although MPC-CasADi attains similar control performance, its computation time is significantly higher due to the absence of ADMM-based acceleration. With the inclusion of MFAPC and Max-Pressure, the advantages of the proposed method become even more pronounced: MFAPC outperforms FTC and MPC-Road by leveraging adaptive pseudogradient updates but remains limited by its model-free structure, while Max-Pressure provides competitive performance under moderate traffic but degrades in oversaturated conditions. In contrast, both FTC and MPC-Road exhibit clearly inferior performance, with higher delays, more stops, and more severe congestion. These observations collectively confirm that the proposed two-lane traffic model, combined with the distributed MPC scheme solved via ADMM, provides substantial improvements over existing fixed-time, model-free, and model-based control methods.

\section{Conclusion}\label{Section V}
In this paper, we proposed a two-lane traffic model for signal control that explicitly captures lane-level vehicle counts and turning movements, addressing the limitations of traditional road-based formulations. By providing a more realistic representation of traffic flow propagation, the model enables improved prediction and control accuracy. Building on this foundation, we developed a distributed MPC framework to optimize phase green times and integrated ADMM into the framework to ensure real-time feasibility when scaling across interconnected intersections. Moreover, the integration of data-driven split-ratio forecasting enhances the controller’s ability to adapt to time-varying turning behaviors, further improving prediction accuracy. Simulation studies conducted on a realistic six-intersection network in Dalian, China, demonstrated the effectiveness of the proposed approach, showing notable improvements in both traffic efficiency and computational performance. A promising avenue for future research is to explore a two-level hierarchical control architecture to further enhance scalability and coordination of the method in large-scale urban traffic networks.

\bibliography{Ting,IDS_Publications_2025,IDS_Publications_12022025}             

@inproceedings{Li2025arouting,
	author = {Li, Anni and Bai, Ting and Chen, Yingqing and Cassandras, Christos G. and Malikopoulos, Andreas A.},
	booktitle = {2025 IEEE 64th Conference on Decision and Control (CDC)},
	pages = {1162-1167},
	title = {A Cooperative Compliance Control Framework for Socially Optimal Mixed Traffic Routing},
	year = {2025}}

@inproceedings{Bai2025routing,
	author = {Bai, Ting and Li, Anni and Xu, Gehui and Cassandras, Christos G. and Malikopoulos, Andreas A.},
	booktitle = {2025 IEEE 64th Conference on Decision and Control (CDC)},
	pages = {1168-1173},
	title = {Routing Guidance for Emerging Transportation Systems with Improved Dynamic Trip Equity},
	year = {2025}}

@article{le2024distributed,
	arxivid = {arXiv:2409.10864},
	author = {Le, Viet-Anh and Malikopoulos, Andreas A},
	journal = {IEEE Control Systems Letters},
	pages = {2721-2726},
	publisher = {IEEE},
	title = {{Distributed optimization for traffic light control and connected automated vehicle coordination in mixed-traffic intersections}},
	volume = {8},
	year = {2024}}

@article{tzortzoglou2024feasibility,
	author = {Tzortzoglou, Filippos N and Beaver, Logan E and Malikopoulos, Andreas A},
	journal = {IEEE Control Systems Letters},
	pages = {2057-2062},
	title = {A feasibility analysis at signal-free intersections},
	volume = {8},
	year = {2024}}

@article{Malikopoulos2020,
	author = {Malikopoulos, Andreas A and Beaver, Logan E and Chremos, Ioannis Vasileios},
	journal = {Automatica},
	number = {109469},
	title = {Optimal Time Trajectory and Coordination for Connected and Automated Vehicles},
	volume = {125},
	year = {2021}}

@article{chalaki2020TITS,
	arxivid = {1911.04082},
	author = {Chalaki, Behdad and Malikopoulos, Andreas A},
	doi = {10.1109/TITS.2021.3123479},
	journal = {IEEE Transactions on Intelligent Transportation Systems},
	number = {8},
	pages = {13330-13345},
	title = {Time-Optimal Coordination for Connected and Automated Vehicles at Adjacent Intersections},
	volume = {23},
	year = {2021}}

@article{chalaki2020TCST,
	arxivid = {2008.02379},
	author = {Chalaki, Behdad and Malikopoulos, Andreas A},
	doi = {10.1109/TCST.2021.3082306},
	journal = {IEEE Transactions on Control Systems Technology},
	number = {3},
	pages = {972--984},
	title = {Optimal Control of Connected and Automated Vehicles at Multiple Adjacent Intersections},
	volume = {30},
	year = {2022}}

@article{ru2023distributed,
  title={Distributed model-free adaptive predictive control of traffic lights for multiple interconnected intersections},
  author={Ru, Xinfeng and Mei, Chenyi and Xia, Weiguo and Shi, Peng},
  journal={SCIENCE CHINA Information Sciences},
  volume={66},
  number={9},
  pages={190209},
  year={2023}
}

@article{yan2016extended,
  title={An extended signal control strategy for urban network traffic flow},
  author={Yan, Fei and Tian, Fuli and Shi, Zhongke},
  journal={Physica A: Statistical Mechanics and Its Applications},
  volume={445},
  pages={117--127},
  year={2016},
  publisher={Elsevier}
}

@ARTICLE{8600793,
  author={Bai, Ting and Li, Shaoyuan and Zheng, Yi},
  journal={IEEE/CAA Journal of Automatica Sinica}, 
  title={Distributed model predictive control for networked plant-wide systems with neighborhood cooperation}, 
  year={2019},
  volume={6},
  number={1},
  pages={108-117},
  }

@article{de2010multi,
  title={Multi-agent model predictive control of signaling split in urban traffic networks},
  author={De Oliveira, Lucas Barcelos and Camponogara, Eduardo},
  journal={Transportation Research Part C: Emerging Technologies},
  volume={18},
  number={1},
  pages={120--139},
  year={2010},
  publisher={Elsevier}
}

@incollection{farokhi2013distributed,
  title={Distributed {MPC} via dual decomposition and alternative direction method of multipliers},
  author={Farokhi, Farhad and Shames, Iman and Johansson, Karl H},
  booktitle={Distributed Model Predictive Control Made Easy},
  pages={115--131},
  year={2013},
  publisher={Springer}
}

@article{bai2020distributed,
  title={Distributed {MPC} for reconfigurable architecture systems via alternating direction method of multipliers},
  author={Bai, Ting and Li, Shaoyuan and Zou, Yuanyuan},
  journal={IEEE/CAA Journal of Automatica Sinica},
  volume={8},
  number={7},
  pages={1336--1344},
  year={2020},
  publisher={IEEE}
}

@article{boyd2011distributed,
  title={Distributed optimization and statistical learning via the alternating direction method of multipliers},
  author={Boyd, Stephen and Parikh, Neal and Chu, Eric and Peleato, Borja and Eckstein, Jonathan and others},
  journal={Foundations and Trends{\textregistered} in Machine learning},
  volume={3},
  number={1},
  pages={1--122},
  year={2011},
  publisher={Now Publishers, Inc.}
}

@Article{Boyd2011,
  author    = {Boyd, Stephen and Parikh, Neal and Chu, Eric and Peleato, Borja and Eckstein, Jonathan and others},
  journal   = {Foundations and Trends{\textregistered} in Machine learning},
  title     = {Distributed optimization and statistical learning via the alternating direction method of multipliers},
  year      = {2011},
  pages     = {1--122},
  volume    = {3},
number    = {1},
  publisher = {Now Publishers, Inc.},
}

@Article{Robertson1969,
  author  = {Robertson, Dennis I},
  journal = {Traffic Engineering \& Control},
  title   = {TRANSYT METHOD FOR AREA TRAFFIC CONTROL},
  year    = {1969},
  number  = {8},
  volume  = {8},
}

@Article{Robertson1991,
  author    = {Robertson, Dennis I and Bretherton, R David},
  journal   = {IEEE Transactions on Vehicular Technology},
  title     = {Optimizing networks of traffic signals in real time--the  {SCOOT} method},
  year      = {1991},
  number    = {1},
  pages     = {11--15},
  volume    = {40},
  publisher = {IEEE},
}

@Article{Sims1980,
  author    = {Sims, Arthur G and Dobinson, Kenneth W},
  journal   = {IEEE Transactions on Vehicular Technology},
  title     = {The {S}ydney coordinated adaptive traffic ({SCAT}) system philosophy and benefits},
  year      = {1980},
  number    = {2},
  pages     = {130--137},
  volume    = {29},
  publisher = {IEEE},
}

@Article{Li2021,
  author    = {Li, Dai and De Schutter, Bart},
  journal   = {IEEE Transactions on Control Systems Technology},
  title     = {Distributed model-free adaptive predictive control for urban traffic networks},
  year      = {2021},
  number    = {1},
  pages     = {180--192},
  volume    = {30},
  publisher = {IEEE},
}

@Book{Hou2013,
  author    = {Hou, Zhongsheng and Jin, Shangtai},
  publisher = {CRC Press Boca Raton, FL},
  title     = {Model Free Adaptive Control},
  year      = {2013},
}

@InProceedings{DeNunzio2015,
  author       = {De Nunzio, Giovanni and Gomes, Gabriel and de Wit, Carlos Canudas and Horowitz, Roberto and Moulin, Philippe},
  booktitle    = {2015 54th IEEE Conference on Decision and Control (CDC)},
  title        = {Arterial bandwidth maximization via signal offsets and variable speed limits control},
  year         = {2015},
  organization = {IEEE},
  pages        = {5142--5148},
}

@inproceedings{ba2015distributed,
  title={Distributed optimal equilibrium selection for traffic flow over networks},
  author={Ba, Qin and Savla, Ketan and Como, Giacomo},
  booktitle={2015 54th IEEE Conference on Decision and Control (CDC)},
  pages={6942--6947},
  year={2015},
  organization={IEEE}
}

@article{bianchin2019,
  title={Gramian-based optimization for the analysis and control of traffic networks},
  author={Bianchin, Gianluca and Pasqualetti, Fabio},
  journal={IEEE Transactions on Intelligent Transportation Systems},
  volume={21},
  number={7},
  pages={3013--3024},
  year={2019},
  publisher={IEEE}
}

@Article{Hou2016,
  author    = {Hou, Zhongsheng and Chi, Ronghu and Gao, Huijun},
  journal   = {IEEE Transactions on Industrial Electronics},
  title     = {An overview of dynamic-linearization-based data-driven control and applications},
  year      = {2016},
  number    = {5},
  pages     = {4076--4090},
  volume    = {64},
  publisher = {IEEE},
}

@article{ye2019survey,
  title={A survey of model predictive control methods for traffic signal control},
  author={Ye, Bao-Lin and Wu, Weimin and Ruan, Keyu and Li, Lingxi and Chen, Tehuan and Gao, Huimin and Chen, Yaobin},
  journal={IEEE/CAA Journal of Automatica Sinica},
  volume={6},
  number={3},
  pages={623--640},
  year={2019},
  publisher={IEEE}
}

@article{zaidi2016back,
  title={Back-pressure traffic signal control with fixed and adaptive routing for urban vehicular networks},
  author={Zaidi, Ali A and Kulcs{\'a}r, Bal{\'a}zs and Wymeersch, Henk},
  journal={IEEE Transactions on Intelligent Transportation Systems},
  volume={17},
  number={8},
  pages={2134--2143},
  year={2016},
  publisher={IEEE}
}

@Article{Xin2011,
  author    = {Lin, Shu and De Schutter, Bart and Xi, Yugeng and Hellendoorn, Hans},
  journal   = {IEEE Transactions on Intelligent Transportation Systems},
  title     = {Fast model predictive control for urban road networks via {MILP}},
  year      = {2011},
  number    = {3},
  pages     = {846--856},
  volume    = {12},
  publisher = {IEEE},
}

@Article{Camponogara2009,
  author    = {Camponogara, Eduardo and De Oliveira, Lucas Barcelos},
  journal   = {IEEE Transactions on Systems, Man, and Cybernetics-Part A: Systems and Humans},
  title     = {Distributed optimization for model predictive control of linear-dynamic networks},
  year      = {2009},
  number    = {6},
  pages     = {1331--1338},
  volume    = {39},
  publisher = {IEEE},
}

@Article{Grandinetti2018,
  author    = {Grandinetti, Pietro and Canudas-de-Wit, Carlos and Garin, Federica},
  journal   = {IEEE Transactions on Control Systems Technology},
  title     = {Distributed optimal traffic lights design for large-scale urban networks},
  year      = {2018},
  number    = {3},
  pages     = {950--963},
  volume    = {27},
  publisher = {IEEE},
}
\end{document}